# Aluminum Scandium Nitride as a Functional Material at 1000°C

*Venkateswarlu Gaddam[1], Shaurya S. Dabas[1], Jinghan Gao[1], David J. Spry[2], Garrett Baucom[1], Nicholas G. Rudawski[1], Tete Yin[1], Ethan Angerhofer[1], Philip G. Neudeck[2], Honggyu Kim[1], Philip X.-L. Feng[1], Mark Sheplak[1], and Roozbeh Tabrizian[1*]*

[1]University of Florida, Gainesville, FL 32611, USA

[2]NASA Glenn Research Center, Cleveland, OH 44135, USA

Keywords: AlScN, extreme environment, ferroelectric, piezoelectric, $d_{33,f}$, electromechanical coupling coefficient, high temperature

Aluminum scandium nitride (AlScN) has emerged as a highly promising material for high-temperature applications due to its robust piezoelectric, ferroelectric, and dielectric properties. This study investigates the behavior of $Al_{0.7}Sc_{0.3}N$ thin films in extreme thermal environments, demonstrating functional stability up to 1000°C, making it suitable for use in aerospace, hypersonics, deep-well, and nuclear reactor systems. Tantalum silicide ($TaSi_2$)/$Al_{0.7}Sc_{0.3}N$/$TaSi_2$ capacitors were fabricated and characterized across a wide temperature range, revealing robust ferroelectric and dielectric properties, along with significant enhancement in piezoelectric performance. At 1000°C, the ferroelectric hysteresis loops showed a substantial reduction in coercive field from 4.3 MV/cm to 1.2 MV/cm, while the longitudinal piezoelectric coefficient increased nearly tenfold, reaching 75.1 pm/V at 800°C. Structural analysis via scanning and transmission electron microscopy confirmed the integrity of the $TaSi_2$/$Al_{0.7}Sc_{0.3}N$ interfaces, even after exposure to extreme temperatures. Furthermore, the electromechanical coupling coefficient was calculated to increase by over 500%, from 12.9% at room temperature to 82% at 700°C. These findings establish AlScN as a versatile material for high-temperature ferroelectric, piezoelectric, and dielectric applications, offering unprecedented thermal stability and functional enhancement.

# 1. Introduction

The development of high-temperature-resilient piezoelectric and ferroelectric thin films is crucial for enabling advanced electronic microsystems to operate reliably in extreme environments, including hypersonic flights, space explorations, deep-well oil and gas operations, and nuclear reactors[1-10].

Conventional ferroelectrics, such as perovskites (*e.g.*, lead-zirconate-titanate) face substantial limitations due to their inability to retain functional stability at temperatures above 200°C, which restricts their use in extreme-temperature applications[5,11]. These limitations have spurred growing interest in alternative materials that can maintain robust functional properties at elevated temperatures, particularly for devices operating between 500°C and 1000°C.

Wurtzite-structured nitride semiconductors, notably aluminum-scandium-nitride (AlScN), have emerged as highly promising candidates for high-temperature applications[1,2,8]. AlScN exhibits exceptional piezoelectric and ferroelectric properties, such as enhanced piezoelectricity, large remanent polarization, and tunable coercive field, while demonstrating superior thermal stability compared to other functional materials[12-24]. Unlike perovskite ferroelectrics, AlScN retains its ferroelectric and piezoelectric behaviors at much higher temperatures, often exceeding 600°C, positioning it as an ideal material for extreme-environment electronics[2,9].

This capability is especially critical for space applications, where electronics that can endure extreme temperatures exceeding 500°C offer significant advancements in systems ranging from hypersonic engines to the scientific explorations of the solar system's harshest environments. For example, NASA has emphasized the need for durable, high-temperature non-volatile memory devices to enable new science missions, such as Venus exploration, where extreme environmental conditions dominate[22]. Given that ferroelectrics have been a key approach to achieving data storage in conventional-temperature non-volatile memory microsystems, the development of resilient ferroelectric storage elements is now being pursued to support durable, extreme-temperature non-volatile semiconductor memory chips[9,23].

High-temperature transducers are equally critical for hypersonic flight vehicle design and gas-turbine technology development. While capacitive and piezoresistive sensors perform well at conventional temperatures, they become inadequate in environments exceeding 500°C, making piezoelectric-based transducers the most viable option for high-sensitivity, high-bandwidth dynamic pressure measurement systems[24,25].

The potential of AlScN extends beyond space explorations and hypersonic systems, with critical implications for industrial applications that require miniaturized, high-performance sensors and systems capable of operating in extreme temperatures. Its ability to maintain strong electromechanical coupling at temperatures over 500°C makes AlScN an attractive candidate for microelectromechanical systems (MEMS) resonators, oscillators, and filters, crucial for radiofrequency (RF) communication, time-keeping, and navigation in harsh environments[26-28]. While initial studies have shown that sputtered AlN films can retain piezoelectric properties up to 1150°C, Sc-alloyed AlN films demonstrate robust piezoelectricity up to at least 600°C, offering promising pathways for the designs of high-temperature resilient MEMS[29].

This paper presents a comprehensive study of the dielectric, piezoelectric, and ferroelectric properties of AlScN across a broad temperature range, from room temperature to 1000°C. We explore the use of a high-temperature-resilient electrode material, tantalum silicide ($TaSi_2$)[30,31], which enables the fabrication of $TaSi_2$/AlScN/$TaSi_2$ capacitors capable of operating at 1000°C. Ferroelectric hysteresis loops were measured from 25°C to 1000°C, revealing a significant reduction in coercive field from 4.3 MV/cm to 1.2 MV/cm. Additionally, longitudinal piezoelectric coefficients ($d_{33,f}$) were measured, showing an unprecedented value of 71.64 pm/V at 1000°C — nearly ten times higher compared to the value at room temperature. Beyond the ferroelectric and piezoelectric properties, various dielectric properties were extracted across this extreme temperature range using current-voltage (*I-V*) and capacitance-voltage (*C-V*) measurements, highlighting the scaling characteristics of these properties.

## 2. Device Fabrication and Characterization Setup

500μm×500μm capacitors were fabricated using a metal-ferroelectric-metal (MFM) stack composed of 700 nm-thick $TaSi_2$ top and bottom electrodes and a 100 nm-thick $Al_{0.7}Sc_{0.3}N$ ferroelectric layer. This MFM stack was deposited onto a silicon substrate covered with a 1 μm-thick silicon dioxide passivation layer.

The $TaSi_2$ films were deposited in an ultrahigh vacuum pulsed-DC sputtering system at 100W under 4 mTorr pressure 30 sccm Kr flow with target-to-substrate spacing of 50 mm and sample lateral motion and rotation[30]. Before depositing the $Al_{0.7}Sc_{0.3}N$ layer, the $TaSi_2$ bottom electrode was annealed at 800°C for 2 hours in a forming gas mixture of 2.5% hydrogen-argon. The $Al_{0.7}Sc_{0.3}N$ layer was deposited at 350°C using a pulsed-DC dual-target Evatec Clusterline 200 II sputter tool. The deposition parameters include a pulse frequency of 150 kHz and a nitrogen flow of 30 sccm. The power applied to the aluminum and scandium targets was set to 1000 W and 545 W, respectively.

X-ray diffractometer (XRD) analysis of the stack showed a full-width-half-maximum (FWHM) of 2.4° for the (002) diffraction peak of $Al_{0.7}Sc_{0.3}N$ at 2$\theta$ of 35.91°. Detailed XRD measurements, both before and after exposure to extreme temperatures, are provided in **Section S1** of the Supporting Document.

The capacitors were patterned by etching the top $TaSi_2$ layer using sulfur hexafluoride in a reactive ion etching tool. The access to the bottom $TaSi_2$ layer was achieved by electrically breaking down sacrificial capacitors with the application of a sufficiently large voltage.

The capacitors were tested in a customized compact high-temperature electrical probing system equipped with a ceramic heater, vacuum chamber (maintained by a dry-scroll pump), cooling system, and precise temperature control. Rhodium microprobes enabled reliable operation up to 1000°C. The system's electrical and optical characterization equipment included a PiezoMEMS Ferroelectric Tester, a Keithly 4200 Semiconductor Analyzer, and a Polytec NLV-2500 Laser Doppler Vibrometer (LDV), allowing for the extraction of ferroelectric, dielectric, conductive, and piezoelectric properties.

**Figure 1a** schematically depicts the sample and the characterization setup. **Figure 1b** presents the temperature profile over time, showing a full cycle from room temperature (25°C) to 1000°C and back to 25°C. **Figure 1c** shows an optical image of the glowing sample at 1000°C and the placement of the LDV for optical interrogation during piezoelectric characterization.

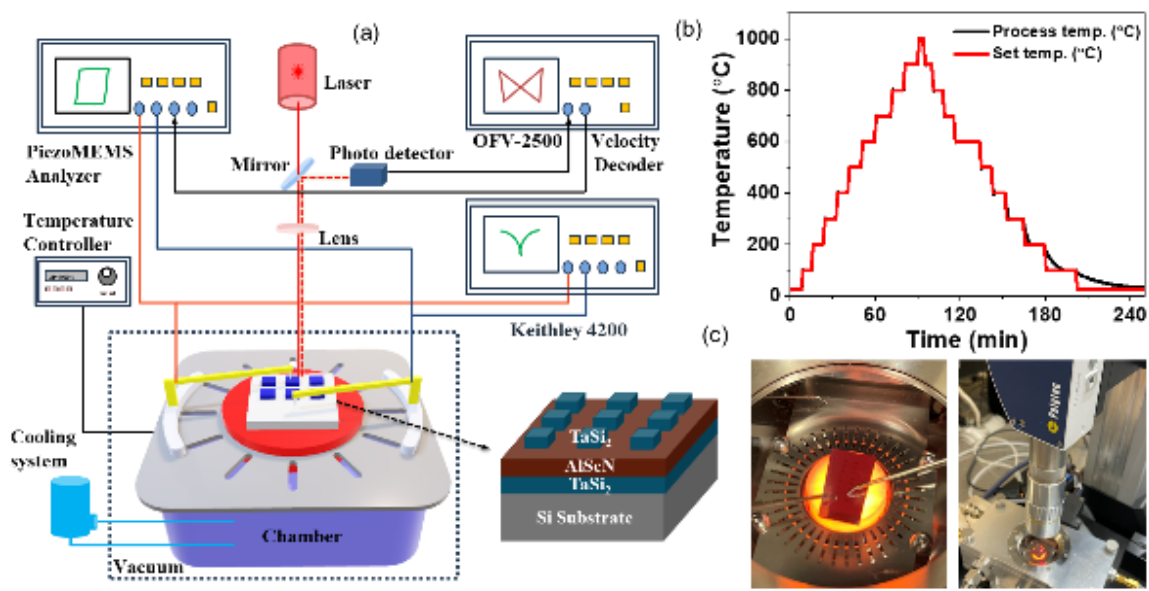

**Figure 1. Experimental Setup for Characterizing AlScN Functional Properties up to 1000°C.** (a) Schematic diagram for the experimental setup of high temperature chamber with electrical and optical feedthroughs enabling extraction of ferroelectric, piezoelectric, dielectric, and conductive properties using $TaSi_2/Al_{0.7}Sc_{0.3}N/TaSi_2$ capacitors. (b) Temperature cycle over time, consisting of raising (25°C to 1000°C) and cooling (1000°C to 25°C) ramps. (c) Optical image of glowing $TaSi_2/Al_{0.7}Sc_{0.3}N/TaSi_2$ capacitors when operating at 1000°C in the vacuum chamber (left) and the placement of LDV atop for piezoelectric characterization (right).

## 3. Structural Analysis using Scanning and Transmission Electron Microscopy

Scanning and Transmission Electron Microscopy (S/TEM) was employed to analyze the structural characteristics of the films and their interfaces. S/TEM sample preparation are provided in the Experimental Section. To assess the impact of high-temperature operation, one capacitor was subjected to a thermal cycle, where the temperature was raised to 1000°C and then cooled back to room temperature, following the profile shown in **Figure 1b**. This thermally cycled capacitor was then used for S/TEM analysis.

**Figure 2a** presents the low-magnification bright-field cross-sectional transmission electron microscopy (BF-XTEM) image of the $TaSi_2/Al_{0.7}Sc_{0.3}N/TaSi_2$ stack. The thicknesses of $Al_{0.7}Sc_{0.3}N$ and $TaSi_2$ layers are approximately 102 nm and 615 nm, respectively, as confirmed by the BF-XTEM images. **Figure 2b** shows scanning transmission electron microscopy (STEM) energy dispersive spectroscopy (EDS) elemental profiles across the $Al_{0.7}Sc_{0.3}N$ layer, superimposed on a high-resolution high-angle annular dark-field imaging STEM (HAADF-STEM) image of the same layer. The EDS profile clearly shows that no detectable amount of electrode material diffused into the $Al_{0.7}Sc_{0.3}N$ film, even after exposure to temperatures as high as 1000°C. Further detailed EDS elemental mapping of the $Al_{0.7}Sc_{0.3}N$ films, shown in Supporting Document **Section S2**, confirms that aluminum, scandium, and nitrogen are uniformly distributed throughout the film, indicating a homogeneous solid solution with no elemental segregation within the limits of the EDS study.

**Figures 2c** and **2e** display high-resolution XTEM (HR-XTEM) images of the $TaSi_2$/AlScN interfaces at the bottom and top of the stack, respectively. **Figure 2d** shows the HR-XTEM image of the central region of the $Al_{0.7}Sc_{0.3}N$ layer, with an inset showing the fast-Fourier-transform (FFT) of the image. The FFT analysis reveals that the $Al_{0.7}Sc_{0.3}N$ layer is distorted single-crystal with strong c-axis orientation, and the zone axis is near B = [11-20]. The d-spacing for the [0002] and [1-100] planes are approximately $0.25 \pm 0.01$ nm and $0.28 \pm 0.01$ nm, respectively, with a c/a ratio of about $1.56 \pm 0.02$. These values are consistent with previous reports.[32]

To investigate the effect of elevated temperatures, a similar S/TEM analysis was conducted on a reference capacitor immediately after fabrication and without thermal treatment. The results, presented in Supporting Document, **Section S3**, showed that the thickness of the $TaSi_2$ electrodes reduced by 7.5% after exposure to 1000°C. Despite the reduction, corresponding to film densification of the $TaSi_2$ electrodes, the HR-XTEM and FFT images of the reference capacitor confirm similar d-spacing ($0.25 \pm 0.01$ nm) and c/a ratio ($1.56 \pm 0.02$) to the annealed sample, suggesting the structural changes due to extreme-temperature exposure are negligible in the AlScN films.

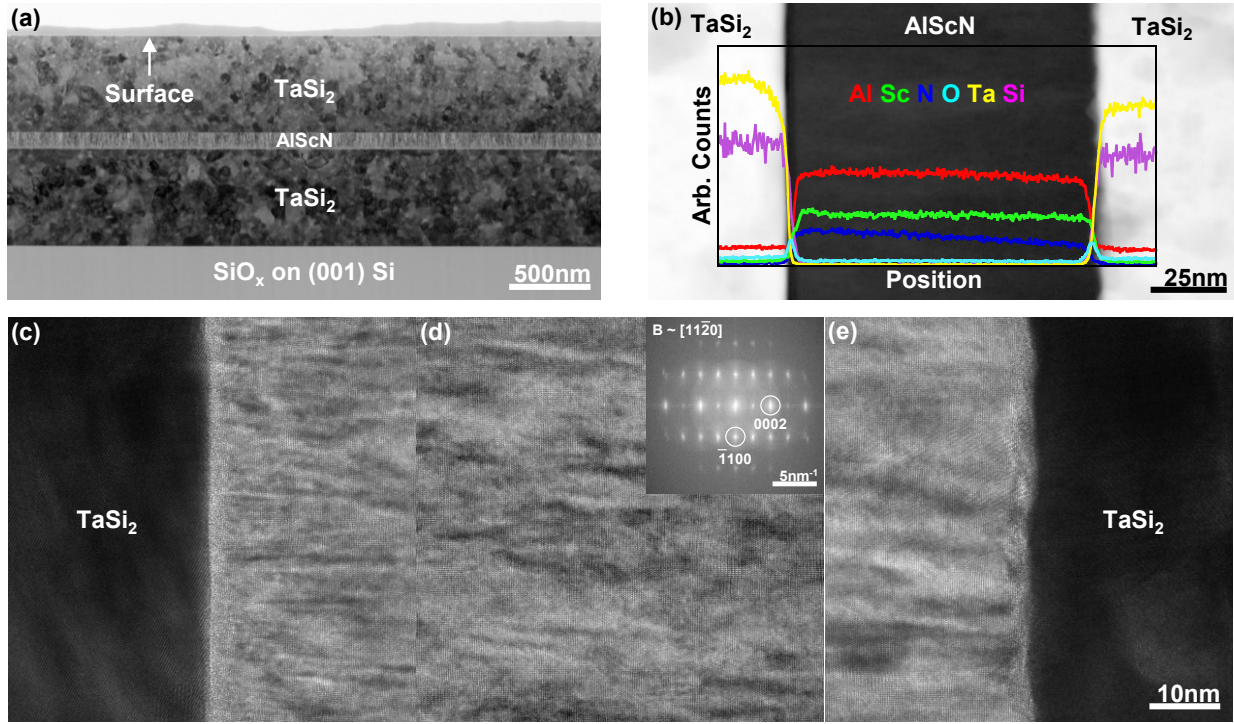

**Figure 2**. **S/TEM characterization of a TaSi$_2$/Al$_{0.7}$Sc$_{0.3}$N/TaSi$_2$ capacitor after extreme temperature cycling.** (a) low magnification BF-XTEM image of the structure; the material on the surface is electron-beam assisted platinum deposited as part of the S/TEM sample preparation process. (b) STEM-EDS elemental profiles across the AlScN layer superimposed on a high-resolution HAADF-STEM image of the AlScN layer. (c) HR-XTEM image of the bottom AlScN/TaSi$_2$ interface. (d) HR-XTEM image of the middle of the AlScN layer (inset: FFT of the image revealing the AlScN layer to be a distorted single crystal with strong c axis orientation). (e) HR-XTEM image of the top AlScN/TaSi$_2$ interface.

## 3. Ferroelectric Characteristics

The ferroelectric polarization hysteresis characteristic of TaSi$_2$/AlScN/TaSi$_2$ capacitors were measured using a Radiant PiezoMEMS Tester. Polarization-electric field (*P-E*) hysteresis loops were captured using a bipolar standard triangular waveform within a frequency range of 31 kHz to 40 kHz. The *P-E* loops were measured over the entire heating and cooling cycle from 25°C to 1000°C to 25°C, following the temperature profile shown in **Figure 1b**. At each temperature, the remanent polarization ($P_r$) and coercive field ($E_c$) were extracted from the loops.

**Figure 3a** shows the measured P-E loops as the temperature increased in 100°C steps, using 35 kHz bipolar triangular drive pulses. **Figure 3b** displays the $E_c$ values extracted from these loops. Similarly, **Figure 3c-d** present the measured *P-E* loops and extracted $E_c$ during cooling, with 31 kHz drive pulses. Across both the heating and cooling half-cycles, a significant change in $E_c$ was observed, decreasing from 4.5 MV/cm at 25°C to 1.2 MV/cm at 1000°C. Both the positive and negative coercive fields (+$E_c$ and -$E_c$) decreased linearly with increasing temperature. This indicates that the elevated thermal energy surpasses the activation barrier for the crystal dipole transformation, facilitating domain switching, consistent with previous studies.[2,9,33] The nearly fourfold reduction in $E_c$ at 1000°C suggests a lower drive voltage is needed for polarization switching at higher temperatures, reducing the need for extreme thickness miniaturization of AlScN for high-temperature memory applications.

The *P-E* loops were also used to extract $P_r$ at different temperatures. **Figure 3e** shows the extracted $P_r$ values during both the heating and cooling ramps. Up to 400°C, $P_r$ remained stable at approximately 83 µC/cm². Over 400°C and 800°C, $P_r$ decreased to about 72 µC/cm². At 1000°C, an apparent increase in $P_r$ was observed, likely due to substantial rise in leakage current. To evaluate the effect of drive voltage frequency on *P-E* loops, additional measurements were conducted at 40 kHz, with results detailed in Supporting Document, **Section S4**.

During the cooling half-cycle, $P_r$ showed monotonic decrease from about 93 µC/cm² to 80 µC/cm² . The variations in $P_r$ between the heating and cooling half-cycles were attributed to the different frequencies of the drive voltages used. To evaluate the effect of the temperature cycling on the *P-E* loop, a recovery hysteresis loop was measured at room temperature after cooling the capacitor down from 1000°C to 25°C, using the same frequency of 35 kHz, as shown in **Figure 3e**. The result indicates that the *P-E* loop was fully recovered, with no significant increase in leakage, demonstrating the resilience of $TaSi_2/AlScN/TaSi_2$ capacitor to high-temperature operation.

To better understand the impact of leakage current on the *P-E* loops at elevated temperatures, the individual hysteresis loops were measured at various temperatures during the cooling half-cycle, at a drive frequency of 31 kHz, as presented in Supporting Document, **Section S5**.

In all measurements, even though a high level of leakage current was evident above 800°C, the polarization switching was still observed, highlighting the potential of AlScN for memory applications at extreme temperatures.

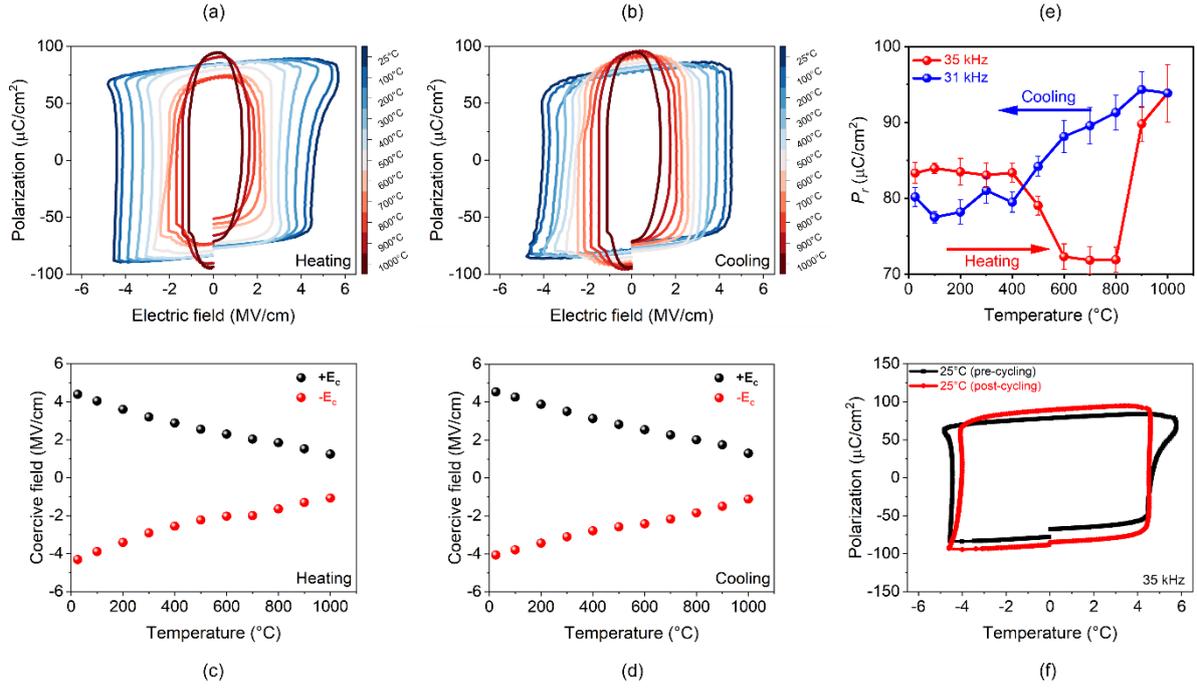

**Figure 3. Ferroelectric characteristics of Al$_{0.7}$Sc$_{0.3}$N over extreme temperature cycling.** Measured P-E curves for the TaSi$_2$/Al$_{0.7}$Sc$_{0.3}$N/TaSi$_2$ capacitors over (a) heating and (b) cooling half-cycles between 25°C to 1000°C. Extracted coercive field values over (c) heating and (d) cooling half-cycles between 25°C to 1000°C. (e) Extracted remanent polarization values for three capacitors, over heating and cooling half-cycles between 25°C to 1000°C. (e) Comparison of P-E loops before and after temperature cycling, measured at 25°C.

## 5. Piezoelectric Characteristics

The effective longitudinal piezoelectric coefficient ($d_{33,f}$) of TaSi$_2$/AlScN/TaSi$_2$ capacitors was measured using a Polytec NLV-2500 LDV system synchronized with a Radiant PiezoMEMS Ferroelectric Tester. Bipolar triangular waveforms at 2 kHz were applied to evaluate the displacement loop. At each temperature, the displacement loops were averaged over five cycles to reduce noise and random drift. Measurements were conducted during both the heating and cooling phases of the thermal cycle, as shown for 4 temperatures in **Figure 4a** and **4b**, respectively. The displacement loops at other temperatures are provided in the Supporting Document, **Section S6**.

The piezoelectric coefficient $d_{33,f}$ was extracted from the linear portion of the displacement loop during voltage reduction to minimize extrinsic contributions, such as domain wall motion. **Figure 4c** presents the $d_{33,f}$ values extracted from displacement loops at different temperatures. At 25°C, a $d_{33,f}$ of 7.6 pm/V was observed. With an increase in temperature to 800°C, the $d_{33,f}$ shows a significant enhancement to 75.1 pm/V. However, further heating to 1000°C led to a decrease in $d_{33,f}$.

This study marks the first observation of such remarkable, nearly tenfold increase in $d_{33,f}$ at high temperatures. However, the underlying mechanisms driving this enhancement require further investigation. We speculate that the enhancement in $d_{33,f}$ may correspond to a change in AlScN crystal phase at extreme temperatures. Similar effects have been reported in perovskite ceramics engineered near polymorphic phase boundaries, where giant piezoelectric coupling occurs[5].

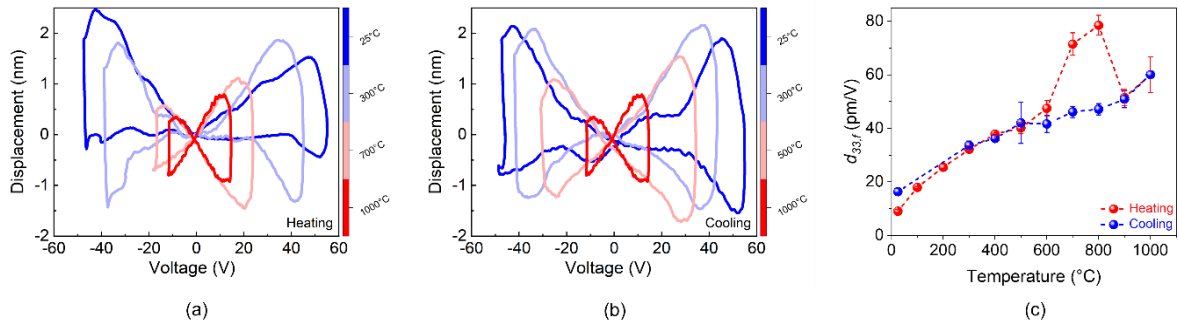

**Figure 4. Piezoelectric characteristics of Al$_{0.7}$Sc$_{0.3}$N over extreme temperature cycling.** Displacement loops for the TaSi$_2$/AlScN/TaSi$_2$ capacitors at different temperatures between 25°C to 1000°C measured over (a) heating and (b) cooling half-cycles. (c) Longitudinal piezoelectric coefficient ($d_{33,f}$) over 25°C to 1000°C, extracted from the displacement loops that were measured for three capacitors, over the extreme temperature cycling.

## 6. Dielectric and Conductive Characteristics

The capacitance-voltage (*C-V*) loops of TaSi$_2$/Al$_{0.7}$Sc$_{0.3}$N/TaSi$_2$ capacitors were measured using the Radiant PiezoMEMS Tester. These *C-V* characteristics were obtained with a switched triangular pulse with drive signal parameters such as tickle voltage of 0.2V, tickle pulse width of 0.001ms, and soak time of 100ms. To determine the optimal frequency, the *C-V* loops were recorded at frequencies ranging from 10 kHz to 2 MHz, with the best results found at 1 MHz. Various voltages were also applied to achieve maximum capacitance with well-defined butterfly-shaped C-V loops. The loops were measured from 25°C to 700°C in 100°C increments, all at a frequency of 1 MHz. At each temperature, the dielectric constant was extracted from the *C-V* loops.

**Figure 5a** presents the *C-V* loop measured at 25°C, showing a butterfly-shaped characteristic of ferroelectric polarization switching, where capacitance decreases as voltage increases. **Figure 5b** show the C-V loop measured at 200°C. As the temperature increases up to 700°C, the capacitance increased, with minimal changes in the loop shape. **Figure 5c** displays the extracted dielectric constants over the temperature range of 25°C to 700°C, showing a consistent increase with temperature. The dielectric constants obtained are comparable to previously reported values.[38] The *C-V* characteristics were not measured beyond 700°C, due to equipment limitations and the large area of the capacitor. Detailed results for different drive frequencies and temperatures are available in the Supporting Document, **Section S7**.

In addition to the large-signal measurements, small-signal capacitance, dielectric constants, and dielectric loss were measured using LCR meter over 25°C to 700°C range, with frequencies of 1 kHz and 10 kHz. **Figure 5d** and **5e** show the measured capacitances and dielectric constants as temperature increased to 700°C, led to a noticeable rise due to an increase in leakage current. **Figure 5f** illustrates the measured dielectric loss up to 700°C, showing a clear increase as temperature rises, which correlates with the large capacitor area and agrees with previous reports.[39,40]

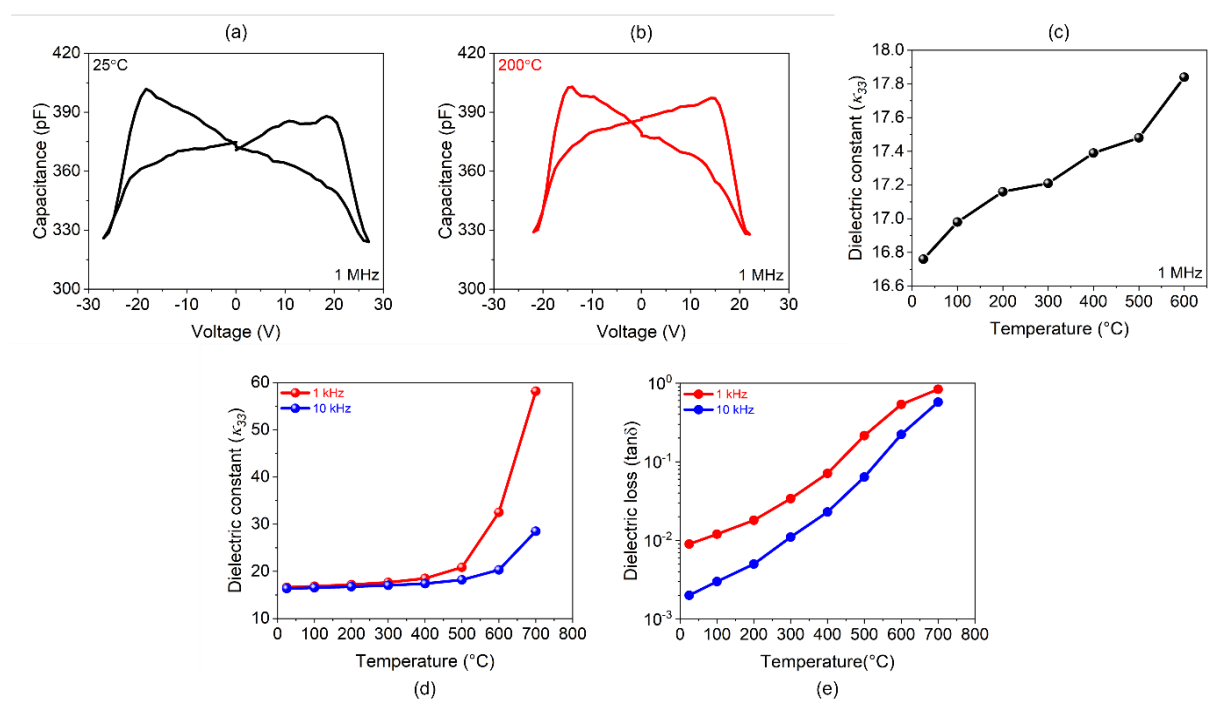

**Figure 5. Dielectric characteristics of Al$_{0.7}$Sc$_{0.3}$N over extreme temperature cycling.** Capacitance-voltage (C-V) loop measured using a 1 MHz drive signal at (a) 25°C and (b) 200°C. (c) Extracted dielectric constants from the measured C-V loops over 25°C to 700°C. (d) Dielectric constant and (e) loss measured by LCR meter at 1 kHz and 10 kHz, over 25°C to 700°C.

To examine conductive properties, current-voltage (*I-V*) characteristics of the TaSi$_2$/Al$_{0.7}$Sc$_{0.3}$N/TaSi$_2$ capacitors were measured using a Keithley 4200 Semiconductor Analyzer. The current density-electric field (*J-E*) characteristics were measured over a temperature range of 25°C to 1000°C, during both heating and cooling half-cycles, as described in **Figure 1b**. At each temperature, J-E characteristics were determined from the *I-V* data using the capacitor's area and AlScN layer thickness. **Figure 6a** shows the J-E characteristics versus temperature in 100°C steps up to 1000°C, where current density rises significantly due to increased leakage current. **Figure 6b** presents the *J-E* characteristics during cooling, showing a reduction in current density as temperature decreases. The large variations in *J* observed during heating and cooling half-cycles are detailed in the Supporting Document, **Section S8**, and may be attributed to thermally activated charge carriers and defects.[41]

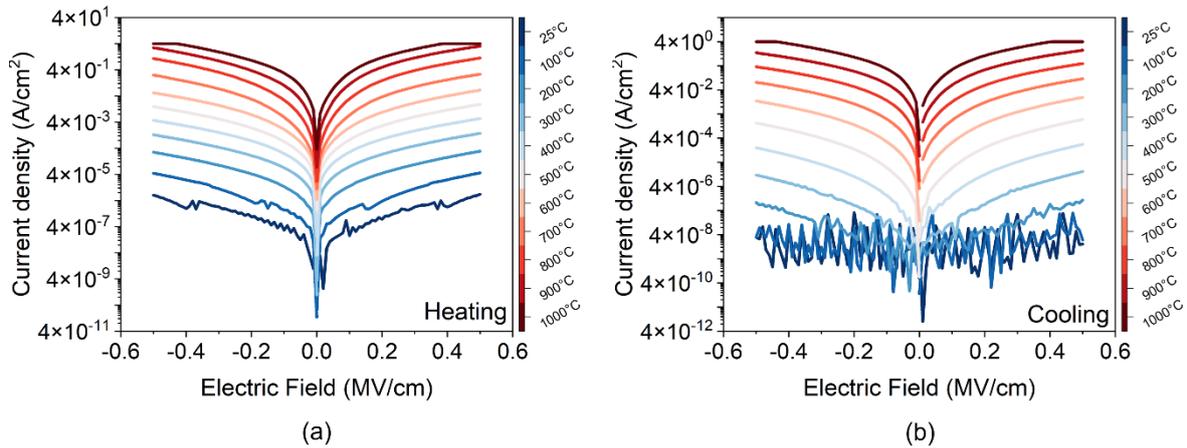

**Figure 6. Conductive characteristics of Al$_{0.7}$Sc$_{0.3}$N over extreme temperature cycling.** Current density versus electric field curves measured over (a) heating and (b) cooling half-cycles between 25°C to 1000°C.

## 7. Benchmark of Ferroelectric and Piezoelectric Characteristics

**Figure 7** compares the measured remanent polarization (*P$_r$*) and the effective longitudinal piezoelectric coefficient (*d$_{33,f}$*) with those of AlScN films in previous reports, over the temperature range of 25°C to 1000°C. As show in in **Figure 7a**, the *P$_r$* measured in this work is lower than the highest values reported for films with similar Sc content. This discrepancy may be due to the differences in crystallinity and texture, which can be influenced by factors such as substrate material, electrode material and thickness, and AlScN film thickness. However, it is important to note that none of the previous studies have demonstrated device survivability beyond 500°C, primarily due to limitations imposed by electrode materials.

**Figure 7b** compares the $d_{33,f}$ values measured in this study with those reported for AlScN films with close composition. While the $d_{33,f}$ values at lower temperatures are comparable with prior reports, the significant enhancement at elevated temperatures sets the results of this work apart. A maximum $d_{33,f}$ of 76 pm/V is measured at 700°C, marking the highest value ever reported for any AlScN film. Using the measured $d_{33,f}$ and dielectric constant, along with reported elasticity for AlScN[37], we estimate the electromechanical coupling ($k_t^2$) across different temperatures. A substantial increase in $k_t^2$ from 12.9% at 25°C to 82% at 700°C is estimated, representing an increase of more than 500%. The detailed calculation procedure of $k_t^2$ is provided in Supporting Document, **Section S9**.

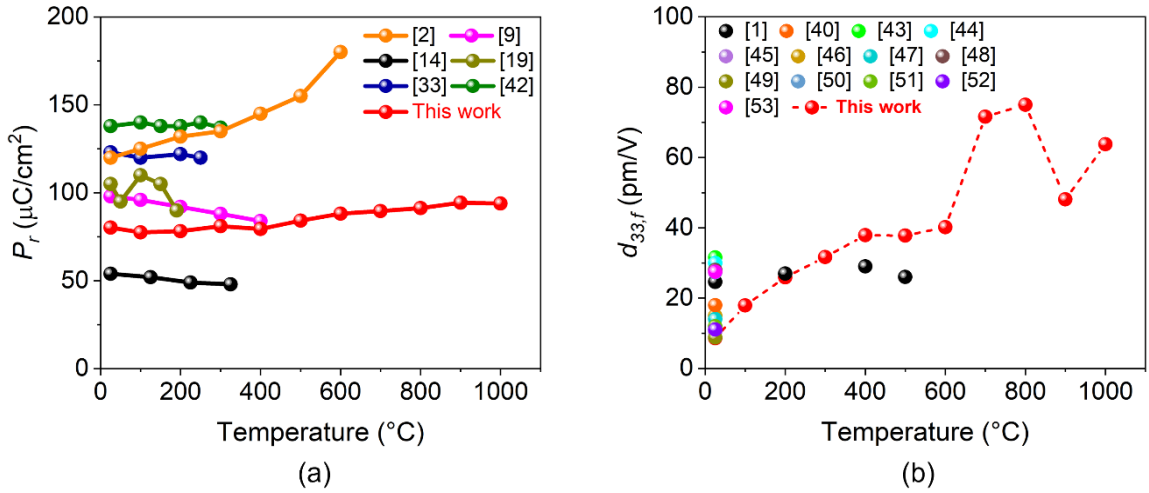

**Figure 7. Comparison of remanent polarization and longitudinal piezoelectric coefficient with previous reports over extreme temperature range.** Comparison of AlScN (a) $P_r$ and (b) $d_{33,f}$ values measured in this work with previous reports, over 25°C to 1000°C.

## 8. Conclusions

This work presents the first comprehensive study of $Al_{0.7}Sc_{0.3}N$-based capacitors operating at extreme temperatures up to 1000°C, highlighting the material's potential for high-temperature electronics. The $TaSi_2/Al_{0.7}Sc_{0.3}N/TaSi_2$ capacitors demonstrated outstanding thermal resilience, maintaining robust piezoelectric, dielectric, and ferroelectric properties, even after cycling through elevated temperatures. Notably, the remanent polarization remained stable and above 72 μC/cm², up to 800°C, while the coercive field exhibited a significant reduction from 4.3 MV/cm to 1.2 MV/cm as the temperature increased to 1000°C. A tenfold increase in the effective longitudinal piezoelectric coefficient at 800°C, and a significant rise in the electromechanical coupling coefficient to 82% at 700°C, further underscore AlScN's superior performance as a transducer for high-temperature environments. This study also highlights the critical role of electrode material selection, with TaSi$_2$ in vacuum characterization environment enabling device survivability beyond 1000°C as well as strong potential for straightforward integration into 500ºC durable SiC integrated circuits process flow[22]. Future work should include further elucidating the mechanisms behind the temperature-induced enhancement in piezoelectric properties, potentially exploring increased Sc content or alternative electrode materials for further performance optimization. These results position AlScN as a leading functional material candidate for extreme-environment microsystems, where reliable operation of piezoelectrics and ferroelectric memory devices at temperatures above 500°C is essential.

## 9. Experimental Section

*Growth of AlScN Thin Films*: A100-nm-thick Al$_{0.7}$Sc$_{0.3}$N layer was deposited using pulsed DC bias (Evatec Clusterline 200 II) sputter tool at 350°C with pulse frequency of 150 kHz, and a gas flow of 30 sccm N$_2$. The Al target and Sc target were kept at 1000W and 545W of power, respectively.

*MFM Capacitors Fabrication*: Prior to growth of AlScN thin films, the bottomTaSi$_2$ electrode was pulse DC sputtered at 100W at 4 mTorr pressure 30 sccm Kr flow with target-to-substrate spacing of 50 mm and sample rotation and lateral motion.[30] The bottom electrode was then annealed at 800°C for 2 hours in a forming gas mixture of 2.5% hydrogen-argon. Then, a 100-nm-thick Al$_{0.7}$Sc$_{0.3}$N layer was deposited using pulsed DC bias sputter tool. Top TaSi$_2$ electrode was deposited over the AlScN thin films using identical TaSi$_2$ sputter deposition procedure as for bottom electrode (without forming gas anneal). Subsequently, the top electrode pattern was defined by standard photolithography process and then etched using deep reactive ion etching (DRIE).

*Materials Characterization:* The thicknesses of the as-deposited AlScN and TaSi$_2$ films were measured by transmission electron microscopy (TEM). Diffraction spectra of AlScN thick films were taken by using X-ray diffractometer (XRD, PANALYTICAL XPERT MRD).

*STEM* imaging and sample preparation:

The capacitors were characterized using scanning/transmission electron microscopy (S/TEM) using a C$_S$ probe-corrected FEI Themis Z scanning/transmission electron microscope operated at 200 kV and equipped with an X-FEG high brightness Schottky field emission gun source, a bottom-mounted Ceta 16 megapixel CMOS camera, Fischione Instruments Model 3000 high-angle annular dark-field scanning transmission electron microscopy (HAADF-STEM) detector, and SuperX windowless SDD energy dispersive spectroscopy (EDS) system. Low magnification bright-field cross-sectional transmission electron microscopy (BF-XTEM) imaging was performed initially to survey the entire capacitor structure followed by high-resolution XTEM (HR-XTEM) and high-resolution HAADF-STEM imaging across the AlScN layer to evaluate the crystal quality and orientation as well as the interfaces with the TaSi$_2$ electrodes. High-resolution STEM-EDS elemental mapping across the AlScN layer was also performed to evaluate the distribution of elements. Bulk lamellas for S/TEM analysis were extracted from the approximate center of the capacitors using an FEI Helios G4 PFIB CXe dual focused ion beam/scanning electron microscope (FIB/SEM) equipped with an EasyLift in-situ micromanipulator using the in-situ lift-out technique[52-54]. However, final thinning of the lamellas to electron transparency was performed using an FEI Helios Nanolab 600i dual FIB/SEM using methods described elsewhere[55-58].

*Electrical Characterization:* Electrical characteristics of the as fabricated TaSi$_2$/AlScN/TaSi$_2$ capacitors such as polarization vs. electric field (*P-E*) hysteresis curves, capacitance vs. voltage (*C-V*) were measured using PiezoMEMs analyzer. Current density vs. electric field (*J–E*) curves were measured using Keithley 4200 parameter analyzer. Displacement vs. voltage curves were measured using Polytec NLV-2500 Laser Doppler Vibrometer and Radiant PiezoMEMS Tester. Further, dielectric loss tangent and small signal capacitance data were measured with LCR meter (TH2811D).

**Supporting Information**

Supporting information consist of extended measurement data over various frequencies, temperatures, and voltages. It also include extensive S/TEM results and electromechanical coupling estimation procedure.

**Acknowledgment**

This work was supported by Defense Advanced Research Projects Agency (DARPA) HOTS program under Grant No. HR00112490309. The authors would like to thank the University of Florida Nanoscale Research Facility cleanroom staff and the NASA Glenn staff, José M. Gonzalez, Carl W. Chang, Srihari Rajgopal, and Gary W. Hunter, for fabrication and administrative supports, and helpful technical discussions.

**Authors' contributions**

V. G., P. N., D. S., P. F., M. S., and R.T. conceived the idea and designed the device fabrication process. D. S. developed and optimized TaSi$_2$ deposition and anneal process. S. D., V. G., and

P. N. fabricated the capacitors. P. N. accomplished all of the TaSi$_2$ runs for the devices of this paper. V. G., J. G., E. A., and T. Y. performed electrical and optical characterizations. N. R., G. B., and H. K. performed S/TEM sample preparation and analysis. M. S., P. F, P. N, H. K, and R. T. supervised the project and provided guidance throughout the process. All authors participated in analyzing the results and contributed to writing the paper. All authors participated in analyzing the results and contributed to writing the paper. All authors have given approval to the final version of the paper.

**Availability of data and materials**

The data that support the findings of this study are available from the corresponding author upon reasonable request.

**Competing interests**

The authors declare that they have no competing interests.

References

1. M. Akiyama, T. Kamohara, K. Kano, A. Teshigahara, Y. Takeuchi, N. Kawahara, **Advanced Materials** 2009, 21, 593.

2. Pradhan, D.K., Moore, D.C., Kim, G. *et al.* A scalable ferroelectric non-volatile memory operating at 600 °C. ***Nat Electron*** **7**, 348–355 (2024).

3. P. French, G. Krijnen, F. Roozeboom, **Microsystem & Nanoengineering** 2016, 2, 16048.

4. S. Zhang, F. Yu, **Journal of the American Ceramic Society** 2011, 94, 3153-3170.

5. P. V. Balachandran, B. Kowalski, A. Sehirlioglu, T. Lookman, **Nature Communications** 2018, 9, 1668.

6. N. A. McDowell, K. S. Knight, P. Lightfoot, **Chemistry – A European Journal** 2006, 12, 1493-1499.

7. T. Stevenson, J. M. Gregg, G. Catalan, **Journal of Materials Science: Materials in Electronics** 2015, 26, 9256-9267.

8. K. Shinekumar, S. Dutta, **Journal of Electronic Materials** 2015, 44, 613-622.

9. D. Drury, M. Kuball, S. Datta, **Micromachines** 2022, 13, 887.

10. M. A. Fraga, F. Furlani, M. Rinaldi, M. Lu, **Microsystem Technologies** 2014, 20, 9-21.

11. I. Stolichnov, A. K. Tagantsev, E. Colla, N. Setter, J. S. Cross, **Journal of Applied Physics** 2005, 98, 84106.

12. S. Fichtner, N. Wolff, F. Lofink, L. Kienle, B. Wagner, **Journal of Applied Physics** 2019, 125, 111101.

13. S. Rassay, F. Hakim, C. Li, C. Forgey, N. Choudhary, R. Tabrizian, **physica status solidi (RRL) – Rapid Research Letters** 2021, 15, 2100087.

14. J. Wang, M. Park, A. Ansari, **2021 IEEE 34th International Conference on Micro Electro Mechanical Systems (MEMS)** 2021, 214-217.

15. P. Wang, D. Wang, S. Mondal, M. Hu, J. Liu, Z. Mi, **Semiconductor Science and Technology** 2023, 38, 043002.

16. Y. Song, C. Perez, G. Esteves, J. S. Lundh, C. B. Saltonstall, T. E. Beechem, J. I. Yang, **ACS Applied Materials & Interfaces** 2021, 13, 19031-19041.

17. D. Wang, J. Zheng, P. Musavigharavi, W. Zhu, A. C. Foucher, S. E. Trolier-McKinstry, E. A. Stach, R. H. Olsson, **IEEE Electron Device Letters** 2020, 41, 1774-1777.

18. D. Q. Tran, F. Tasnádi, A. Žukauskaitė, J. Birch, V. Darakchieva, P. P. Paskov, **Applied Physics Letters** 2023, 122, 181903.

19. V. Gund, B. Davaji, H. Lee, M. J. Asadi, J. Casamento, H. G. Xing, D. Jena, A. Lal, **IEEE International Symposium on Applications of Ferroelectric, ISAF 2021**.

20. N. Wolff, M. R. Islam, L. Kirste, S. Fichtner, F. Lofink, A. Žukauskaitė, L. Kienle, **Micromachines** 2022, 13, 1282.

21. K. H. Kim, I. Karpov, R. H. Olsson, D. Jariwala, **Nature Nanotechnology** 2023, 18, 422.

22. T. Kremic, M. Amato, M. S. Gilmore, W. S. Kiefer, N. M. Johnson, J. Sauder, G. W. Hunter, T. Thompson, **Venus Surface Platform Study Final Report**, Glenn Research Center, 2021.

23. Y. He, S. Chen, M. M. A. Fiagbenu, C. Leblanc, P. Musavigharavi, G. Kim, X. Du, **Applied Physics Letters** 2023, 123, 123504.

24. J. Samuels, S. G. Roberts, **Proceedings of the Royal Society A** 1989, 421, 1-23.

25. D.A., Mills, W. Patterson, P. Fournier, D. Trabbic, J. Underbrink, M. Sheplak, **In** *30th AIAA/CEAS Aeroacoustics Conference* 2024, p.3061.

26. D. Mo, S. Dabas, S. Rassay, R. Tabrizian, **IEEE Transactions on Electron Devices** 2022, 69, 4624-4631.

27. S. Dabas, D. Mo, B. Chatterjee, R. Tabrizian, **2024 IEEE 37th International Conference on Micro Electro Mechanical Systems (MEMS)** 2024, 152-155.

28. Q. Wang, Y. Lu, S. Fung, X. Jiang, S. Mishin, Y. Oshmyansky, D. A. Horsley, **Proceedings of the Solid-State Sensors, Actuators, and Microsystems Workshop**, 2016, 436-439.

29. W. Sui, P. X.-L. Feng, **Applied Physics Letters** 2024, 125, 022201.

30. D. J. Spry, P. G. Neudeck, C. W. Chang, S. Rajgopal, J. M. Gonzalez, **Key Engineering Materials** 2023, 948, 83-88.

31. A. V. Nomoev, S. P. Bardakhanov, M. Schreiber, D. Z. Bazarova, B. B. Baldanov, N. A. Romanov, **Nanomaterials** 2014, 5, 26-35.

32. M. R. Islam, N. Wolff, M. Yassine, G. Schönweger, B. Christian, H. Kohlstedt, O. Ambacher, F. Lofink, L. Kienle, S. Fichtner, **Appl. Phys. Lett.** 2021, 118, 232905.

33. W. Zhu, J. Hayden, F. He, J. I. Yang, P. Tipsawat, M. D. Hossain, J. P. Maria, S. Trolier-McKinstry, *Appl Phys Lett* **2021**, *119*, 62901.

34. Meier, D. & Selbach, S. M. Ferroelectric domain walls for nanotechnology. Nat. Rev. Mater. 7, 157–173 (2022).

35. K. Kim, Y.B. Lee, S.H. Lee, I. S. Lee, S. K. Ryoo, S. Y. Byun, J. H. Lee, C. S. Hwang, Impact of Operation Voltage and $NH_3$ Annealing on the Fatigue Characteristics of Ferroelectric AlScN Thin Films Grown by Sputtering. *Nanoscale* **2023**, *15*, 16390–16402

36. J. Wang, M. Park, A. Ansari, High-Temperature Acoustic and Electric Characterization of Ferroelectric $Al_{0.7}Sc_{0.3}N$ Films, **Journal of Microelectromechanical Systems** **2022**, 8;31(2):234-240

37. Y. Dai, S. Li, H. Gao, W. Wang, Q. Sun, Q. Peng, C. Gui, Z. Qian, S. Liu, Stress evolution in AlN and GaN grown on Si (111): experiments and theoretical modeling. **Journal of Materials Science: Materials in Electronics**, 2016, 27, pp.2004-2013

38. S. Bals, W. Tirry, R. Geurts, Y. Zhiqing, D. Schryvers, *Microscopy and Microanalysis* **2007**, *13*, 80.

39. S. Mueller, J. Müller, U. Schroeder, T. Mikolajick, *IEEE transactions on device and materials reliability* **2013**, *13*, 93.

40. M. Pirro, B. Herrera, M. Assylbekova, G. Giribaldi, L. Colombo, M. Rinaldi, Characterization of dielectric and piezoelectric properties of ferroelectric AlScN thin films, *in Proc. IEEE 34th Int. Conf. Micro Electro Mech. Syst. (MEMS),* **2021**, pp. 646–649.

41. R. Guido, P.D. Lomenzo, M.R. Islam, N. Wolff, M. Gremmel, G. Schönweger, H. Kohlstedt, L. Kienle, T. Mikolajick, S. Fichtner, U. Schroeder, Thermal stability of the ferroelectric properties in 100 nm-thick $Al_{0.72}Sc_{0.28}N$. *ACS Applied Materials & Interfaces*, **2023**, *15*(5), pp.7030-7043.

42. R. Mizutani, S. Yasuoka, T. Shiraishi, T. Shimizu, M. Uehara, H. Yamada, M. Akiyama, O. Sakata, H. Funakubo, Thickness scaling of $(Al_{0.8}Sc_{0.2})N$ films with remanent polarization beyond 100 μC $cm^{-2}$ around 10 nm in thickness. *Applied Physics Express*, **2021**, *14*(10), p.105501.

43. Y. Lu, M. Reusch, N. Kurz, A. Ding, T. Christoph, M. Prescher, L. Kirste, O. Ambacher, A. Žukauskaitė, Elastic modulus and coefficient of thermal expansion of piezoelectric $Al_{1-x}Sc_xN$ (up to x= 0.41) thin films, *Apl Materials*, **2018**, *6*(7), p. 076105.

44. S. Barth, H. Bartzsch, D. Gloess, P. Frach, T. Herzog, S. Walter, H. Heuer, Sputter deposition of stress-controlled piezoelectric AlN and AlScN films for ultrasonic and energy harvesting applications. *IEEE transactions on ultrasonics, ferroelectrics, and frequency control*, *61*(8), **2014**, pp.1329-1334.

45. P.M. Mayrhofer, E. Wistrela, M. Schneider, A. Bittner, u. Schmid, U., Precise determination of d33 and d31 from piezoelectric deflection measurements and 2D FEM simulations applied to $Sc_xAl_{1-x}N$. *Procedia Engineering*, **2016**, *168*, pp.876-879

46. J. Casamento, C.S. Chang, Y.T. Shao, J. Wright, D.A. Muller, H.G. Xing, D. Jena, Structural and piezoelectric properties of ultra-thin $Sc_xAl_{1-x}N$ films grown on GaN by molecular beam epitaxy. *Applied Physics Letters*, 2020, *117*(11), p. 112101.

47. Y. Kusano, G.L. Luo, D. Horsley, I. Taru Ishii, A. Teshigahara, 36% scandium-doped aluminum nitride piezoelectric micromachined ultrasonic transducers. In *2018 IEEE International Ultrasonics Symposium (IUS), IEEE,* **2018**, pp. 1-4.

48. K. Umeda, H. Kawai, A. Honda, M. Akiyama, T. Kato, T. Fukura, Piezoelectric properties of ScAlN thin films for piezo-MEMS devices. In *2013 IEEE 26th International Conference on Micro Electro Mechanical Systems (MEMS), IEEE,* **2013**, pp. 733-736.

49. J. Patidar, K. Thorwarth, T. Schmitz-Kempen, R. Kessels, S. Siol, Deposition of highly crystalline AlScN thin films using synchronized high-power impulse magnetron sputtering: From combinatorial screening to piezoelectric devices. *Physical Review Materials*, **2024,** *8*(9), p.095001.

50. O. Zywitzki, T. Modes, S. Barth, H. Bartzsch, P. Frach, Effect of scandium content on structure and piezoelectric properties of AlScN films deposited by reactive pulse magnetron sputtering. *Surface and Coatings Technology*, **2017**, *309*, pp.417-422.

51. M.A. Signore, A. Serra, D. Manno, G. Quarta, L. Calcagnile, L. Maruccio, E. Sciurti, E. Melissano, A. Campa, M.C. Martucci, L. Francioso, Low temperature sputtering deposition of Al1−xScxN thin films: Physical, chemical, and piezoelectric properties evolution by tuning the nitrogen flux in (Ar+ N2) reactive atmosphere. *Journal of Applied Physics*, **2024**, *135*(12), p. 125105.

52. L.A. Giannuzzi, F.A Stevie, A review of focused ion beam milling techniques for TEM specimen preparation. *Micron*, **1999**, *30*(3), pp.197-204.

53. L.A. Giannuzzi, J. L. Drown, S.R. Brown, R. B. Irwin, F.A. Stevie, Focused ion beam milling and micromanipulation lift-out for site specific cross-section TEM specimen preparation. *MRS Online Proceedings Library*, **1997**, *480*(1), pp.19-27.

54. F.A Stevie, C.B. Vartuli, L.A. Giannuzzi, T.L. Shofner, S.R. Brown, B. Rossie, F. Hillion, R.H. Mills, M. Antonell, R.B. Irwin, B.M. Purcell, Application of focused ion beam lift-out specimen preparation to TEM, SEM, STEM, AES and SIMS analysis. *Surface and Interface Analysis: An International Journal devoted to the development and application of techniques for the analysis of surfaces, interfaces and thin films*, **2001**, *31*(5), pp.345-351.

55. M. Schaffer, B. Schaffer, Q. Ramasse, Sample preparation for atomic-resolution STEM at low voltages by FIB. *Ultramicroscopy*, **2012**, *114*, pp.62-71.

56. J. Mayer, L.A. Giannuzzi, T. Kamino, J. Michael, TEM sample preparation and FIB-induced damage. *MRS bulletin*, **2007**, *32*(5), pp.400-407.

57. L.A. Giannuzzi, Reducing FIB damage using low energy ions. *Microscopy and Microanalysis*, **2006**, *12*(S02), pp.1260-1261.

58. S. Bals, W. Tirry, R. Geurts, Z. Yang, D. Schryvers, High-quality sample preparation by low kV FIB thinning for analytical TEM measurements. *Microscopy and Microanalysis*, **2007**, *13*(2), pp.80-86.

**Title: "AlScN Thin Films Enabling Microsystems at Extreme Temperatures"**

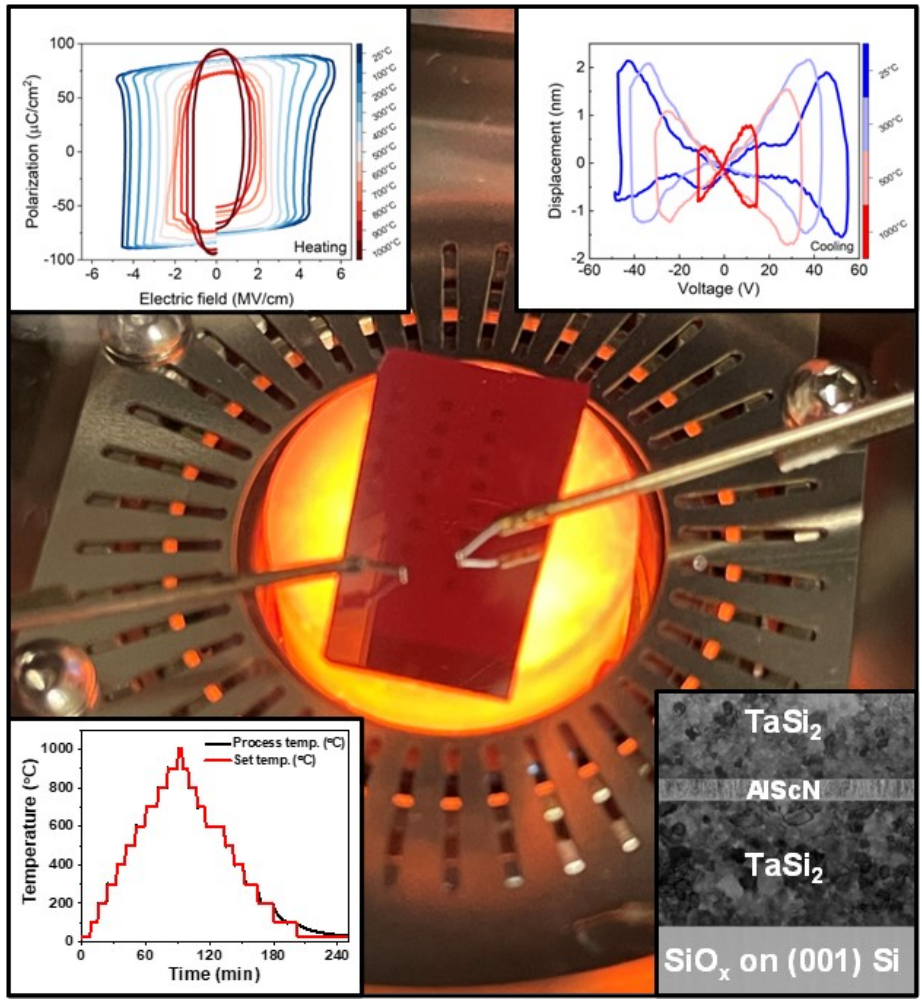